\documentstyle[twocolumn,prl,aps,epsfig,amssymb]{revtex}
\tighten
\let\jnfont=\rm
\def\NPB#1,{{\jnfont Nucl.\ Phys.\ }{\bf B#1},}
\def\PLB#1,{{\jnfont Phys.\ Lett.\ B }{\bf #1},}
\def\PRD#1,{{\jnfont Phys.\ Rev.\ D }{\bf #1},}
\def\PRL#1,{{\jnfont Phys.\ Rev.\ Lett.\ }{\bf #1},}
\def\ZPC#1,{{\jnfont Z.\ Phys.\ C }{\bf #1},}
\def\EPJC#1,{{\jnfont Eur.\ Phys.\ J.\ C }{\bf #1},}
\begin{document}
\draft
\preprint{}

\title{ \hfill {\small TU-767} \\
        \hfill {\small hep-ph/0604163}\\  ~~\\
       Experimental Constraints on Scharm-Stop Flavor Mixing \\
            and Implications in Top-quark FCNC Processes}

\author{Junjie Cao$^1$,  Gad Eilam$^1$, Ken-ichi Hikasa$^2$, Jin Min Yang$^3$ }

\address{ \ \\[2mm]
{\it $^1$ Physics Department, Technion, 32000 Haifa, Israel}\\[2mm]
{\it $^2$ Department of Physics, Tohoku University, Sendai 980-8578, Japan}\\[2mm]
{\it $^3$ Institute of Theoretical Physics, Academia Sinica, Beijing 100080, China}\\[6mm] }

\maketitle

\begin{abstract}
We examine experimental constraints on scharm-stop flavor mixing
in the minimal supersymmetric standard model, which arise from the
experimental bounds on squark and Higgs boson masses, the
precision measurements of $W$-boson mass and the effective weak
mixing angle, as well as the experimental data on $B_s-\bar{B}_s$
mixing and $b \to s \gamma$. We find that the combined analysis
can put rather stringent constraints on $\tilde{c}_L-\tilde{t}_L$
and $\tilde{c}_L-\tilde{t}_R$ mixings. As an illustration for the
effects of such constraints, we examine various top-quark
flavor-changing neutral-current processes induced by scharm-stop
mixings at the LHC and find that their maximal rates are
significantly lowered.
\end{abstract}

\pacs{14.80.Ly, 11.30.Hv}

\bf{Introduction~~} \rm It is well known that in the minimal
supersymmetric standard model (MSSM), besides CKM matrix, flavor
mixings in sfermion sector are another source of flavor violation
\cite{susyf}. Since such mixings arise from soft breaking terms,
they relate flavor problem to SUSY breaking and their information
may provide guidelines for SUSY model building. While such mixings
can be directly measured through the flavor-changing decays of
sfermions at future colliders \cite{taohan}, their information can
also be obtained from various low energy processes
\cite{susyf,bsr}. So far the flavor mixings involving the
first-generation squarks have been severely constrained by
$K^0-\bar K^0$, $D^0-\bar D^0$ and $B^0_d-\bar B^0_d$ mixings
\cite{susyf}, but the mixings between second- and third-generation
squarks, especially the scharm-stop mixings, are less constrained.
For example, although the electric dipole moment of mercury atom
has been precisely measured, it only constrains the product of the
$\tilde{c}_L-\tilde{t}_L$ mixing with another undetermined free
parameter \cite{Endo}. Without definite information about the
parameter, the constraint is ambiguous.

Scharm-stop mixings are well motivated in popular flavor-blind
SUSY breaking models like supergravity models (SUGRA). In these
models, the flavor-diagonality is usually assumed in sfermion mass
matrices at the grand unification scale and the Yukawa couplings
induce flavor mixings when the sfermion mass matrices evolve down
to the weak scale. Such radiatively induced off-diagonal
squark-mass terms are proportional to the corresponding Yukawa
couplings and thus the mixings between second- and
third-generation squarks may be sizable\cite{hikasa}.

Scharm-stop mixings can induce various top-quark flavor-changing
neutral-current (FCNC) processes which will be tested at the CERN
Large Hadron Collider (LHC). Given the importance of such mixings,
we in this work examine current experimental constraints on them
from the following considerations. Firstly, since the mixing terms
appear as the non-diagonal elements of squark mass matrices, they
can affect the squark mass spectrum, especially enlarge the mass
splitting between squarks. So they should be constrained by the
squark mass bounds from the direct experimental searches. At the
same time, since the squark loops affect the precision electroweak
quantities such as $M_W$ and the effective weak mixing angle
$sin^2\theta_{eff}$, such mixings could be also constrained by the
precision electroweak measurements. As will be shown later, to a
good approximation, the supersymmetric corrections to the
electroweak quantities are through the parameter $\delta\rho$ and
thus sensitive to the mass splitting of squarks. Secondly, the
processes of $b \to s$ transition like $B_s-\bar B_s$ mixings and
$b \to s \gamma$ can provide rich information about the $\tilde
s-\tilde b$ mixings. Through the SU(2) relation between up-squark
and down-squark mass matrices (see eq.(\ref{SU2})) and also
through the electroweak quantities (since all squarks contribute
to electroweak quantities via loops), the information can be
reflected in up-squark sector and hence constrain the scharm-stop
mixings. Thirdly, we note that the chiral flipping mixings of
scharm-stop come from the trilinear $H_2 \tilde{Q} \tilde{U}$
interactions \cite{susyf}. Such interactions can lower the
lightest Higgs boson mass $m_h$ via squark loops and thus should
be subject to the current experimental bound on $m_h$.

We noticed that the constraints on scharm-stop mixings from $m_h$
and $\delta \rho$ have been discussed in \cite{hollik}. But the
analyses of \cite{hollik} focus on $\tilde c_L-\tilde t_L$ mixing
and did not consider $\tilde c_L-\tilde t_R$, $\tilde c_R-\tilde
t_L$ and $\tilde c_R-\tilde t_R$ mixings. Also,  in \cite{hollik}
the authors ignored the SU(2) relation between the up- and
down-squark mass matrices, and for the up-squark mass matrix, they
left out the first generation and only considered the second and
third generations. As will be discussed later, a complete
consideration of three generations is necessary in calculating
$\delta \rho$.  Moreover, since the analyses of \cite{hollik}
focus on $m_h$ and $\delta \rho$ constraints, other constraints
like $B_s-\bar{B}_s$ mixing and $ b \to s \gamma$ are not included
in \cite{hollik}.

In this work we will give a comprehensive analysis on the
scharm-stop mixings. We will consider all possible mixings between
scharms and stops, namely $\tilde c_L-\tilde t_L$ and $\tilde
c_L-\tilde t_R$ mixings for left-handed scharm and $\tilde
c_R-\tilde t_L$ and $\tilde c_R-\tilde t_R$ mixings for
right-handed scharm. We will not only consider the constraints
from $m_h$ and $\delta \rho$, but also include the constraints
from $B_s-\bar{B}_s$ mixing and $b \to s \gamma$. As an
illustration for the effects of such constraints, we will examine
various top-quark FCNC processes induced by these scharm-stop
mixings at the LHC. \vspace*{0.2cm}

\bf{Calculations~~}  \rm Instead of presenting the detailed and lengthy
analytic results, we just delineate the strategies of our
calculations. \vspace*{0.1cm}

\em{1. Squark mass:~~} \rm In the super-KM basis with states
($\tilde u_L$, $\tilde c_L$, $\tilde t_L$, $\tilde u_R$, $\tilde
c_R$, $\tilde t_R$) for up-squarks and  ($\tilde d_L$, $\tilde
s_L$, $\tilde b_L$, $\tilde d_R$, $\tilde s_R$, $\tilde b_R$) for
down-squarks, the $6\times 6$ squark mass matrix
${\cal M}^2_{\tilde q}$ ($\tilde q=\tilde u, \tilde d$) takes the
form \cite{susyf}
\begin{eqnarray}
{\cal M}^2_{\tilde q}=\left( \begin{array}{ll}
  (M^2_{\tilde q})_{LL}+ C_{\tilde q}^{LL} & (M^2_{\tilde q})_{LR}-C_{\tilde q}^{LR} \\
  \left( (M^2_{\tilde{q}})_{LR}-C_{\tilde q}^{LR}\right)^\dag
             & (M^2_{\tilde{q}})_{RR}+C_{\tilde q}^{RR} \end{array} \right) ,
\label{sq-matrix}
\end{eqnarray}
where $C_{\tilde q}^{LL} = m_q^2 + \cos 2\beta M_Z^2 ( T_3^q - Q_q
s_W^2) \hat{\mbox{\large 1}}$, $C_{\tilde q}^{RR} = m_q^2 + \cos
2\beta M_Z^2 Q_q s_W^2 \hat{\mbox{\large 1}}$ and $C_{\tilde
q}^{LR}=m_q \mu (\tan \beta)^{- 2 T_3^q}$ are $3\times 3$ diagonal
matrices ($\hat{\mbox{\large 1}}$ stands for the unit matrix in
flavor space and $m_q$ is the diagonal quark mass matrix). Here,
$T_3^q=1/2$ for up-squarks and  $T_3^q=-1/2$ for down-squarks, and
$\tan \beta =v_2/v_1$ is the ratio of the vacuum expectation
values of the Higgs fields. In general, the soft mass parameters
$(M^2_{\tilde q})_{LL}$,  $(M^2_{\tilde q})_{LR}$ and
$(M^2_{\tilde q})_{RR}$ are $3\times 3$ non-diagonal matrices. For
up-squarks, if we only consider the flavor mixings between scharms
and stops, then
\begin{eqnarray}
(M^2_{\tilde{u}})_{LL} &= & \left ( \begin{array}{ccc}
    M_{Q_1}^2 &  0                           &  0 \\
    0         &  M_{Q_2}^2                   & \delta_{LL} M_{Q_2} M_{Q_3} \\
    0         &  \delta_{LL} M_{Q_2} M_{Q_3} & M_{Q_3}^2 \end{array}  \right ), \nonumber \\
 (M^2_{\tilde{u}})_{LR} &=&  \left (
\begin{array}{ccc}
0     &  0   &  0 \\
0     &  0   & \delta_{LR} M_{Q_2} M_{U_3}\\
0     &  \delta_{RL} M_{U_2} M_{Q_3}  & m_t A_t \end{array}  \right ), \nonumber \\
 (M^2_{\tilde{u}})_{RR} &= & (M^2_{\tilde{u}})_{LL}|_{M_{Q_i}^2 \to M_{U_i}^2,~ \delta_{LL} \to
 \delta_{RR}}. \label{up squark}
\end{eqnarray}
Similarly, for down-squarks we have
\begin{eqnarray}
(M^2_{\tilde{d}})_{LR}& = & \left ( \begin{array}{ccc}
0          &  0                  &  0 \\
0                                &  0   & \delta_{LR}^d M_{Q_2} M_{D_3}\\
0 &  \delta_{RL}^d M_{D_2} M_{Q_3}  & m_b A_b \end{array}  \right
),  \nonumber  \\
(M^2_{\tilde{d}})_{RR} &= & (M^2_{\tilde{u}})_{LL}|_{M_{Q_i}^2 \to
M_{D_i}^2,~ \delta_{LL} \to \delta_{RR}^d}, \label{delta'-LR}
\end{eqnarray}
and, due to $SU_L(2)$ gauge invariance, $(M^2_{\tilde d })_{LL}$
is determined by\cite{susyf}
\begin{eqnarray}
(M^2_{\tilde{d}})_{LL} = V_{CKM}^\dag (M_{\tilde{u}}^2)_{LL}
V_{CKM}.  \label{SU2}
\end{eqnarray}
Note that for the diagonal elements of left-right mixings in
eqs.(\ref{up squark}$\sim$\ref{delta'-LR}), we only kept the terms
of third-family squarks since we adopted the popular assumption
that they are proportional to the corresponding quark masses.

The squark mass eigenstates can be obtained by diagonalizing the
mass matrix in eq.(\ref{sq-matrix}) with an unitary rotation
$U_{\tilde{q}}$, which is performed numerically in our analysis.
The interactions of a vector or scalar boson $X$ with a pair of
squark mass eigenstates are then obtained by
\begin{eqnarray} V(X \tilde{q}_{\alpha}^\ast
\tilde{q}_{\beta}^\prime) \; = \;
 U_{\tilde{q}}^{\dag \alpha,i} \;  U_{\tilde{q}^\prime}^{j, \beta} \;
 V(X \tilde{q}_i^\ast \tilde{q}^\prime_j)~,
\label{rotation}
\end{eqnarray}
where $V(X\tilde{q}_i^\ast \tilde{q}^\prime_j)$ denotes a generic
vertex in the interaction basis and
$V(X\tilde{q}_{\alpha}^\ast \tilde{q}_{\beta}^\prime )$ is the
vertex in the mass-eigenstate basis. It is clear that both the
squark masses and their interactions depend on the mixing
parameters in the squark mass matrices.

Although we in eq.(\ref{up squark}) listed all four possible
mixings between scharms and stops, in the following we mainly
focus on the mixings $\delta_{LL}$ and $ \delta_{LR}$ for the
left-handed scharm, and only give some brief discussions about the
mixings $\delta_{RL} $ and $\delta_{RR}$ for the right-handed
scharm. Our peculiar interest in $\delta_{LL}$ and $\delta_{LR}$
is based on the following two considerations. The first is that in
the popular mSUGRA model, at the weak scale the flavor mixings for
the left-handed scharm are proportional to bottom quark mass while
those for the right-handed scharm are proportional to charm quark
mass\cite{hikasa}. Therefore, in phenomenological studies of
scharm-stop mixings, one usually assume the existence of
$\delta_{LL}$ and $\delta_{LR}$. The second is that $\delta_{LL}$
and $\delta_{LR}$ have richer phenomenology than $\delta_{RL}$ and
$\delta_{RR}$. $\delta_{LL}$ and $\delta_{LR}$ contribute sizably
to all considered quantities, namely $\delta \rho$, $m_h$,
$B_s-\bar{B}_s$ mixing and $ b \to s \gamma$, while $\delta_{RL}$
and $\delta_{RR}$ only affect $\delta \rho $ and $m_h$.
\vspace*{0.2cm}

\em{2. $b \to s \gamma$:~~} \rm It has long been known that
for $b \to s \gamma$ the sizable SUSY contributions may come
from three kinds of loops mediated respectively by
the charged Higgs bosons, charginos and gluinos \cite{charged higgs}.
In our analysis we consider all these three kinds of loops plus
those mediated by the neutralinos.
We use the formula in \cite{bsr}, which includes all these
SUSY loop effects in additional to the SM contribution.
So, besides down-squark and up-squark mass parameters, our results
also depend on charged Higgs boson mass, gaugino mass $M_2$ and gluino mass
$m_{\tilde{g}}$. As pointed out in numerous papers, $b \to s
\gamma$ is very sensitive to $\delta^d_{LR}$ and  $\delta^d_{RL}$,
and in some cases also sensitive to $\delta_{LL}$ and
$\delta_{LR}$.

\em{3. $B_s-\bar{B}_s$ mixing:~~} \rm In the MSSM the charged
Higgs and chargino contributions to $B_s-\bar{B}_s$ mixing are
much suppressed compared with the gluino effects and the SM
prediction \cite{Ball}. So in our analysis we only include the
gluino effects in addition to the SM contribution. We evaluate the
gluino contributions by the full expressions, namely, without the
mass-insertion approximation. Our Wilson coefficients are coincide
with those in \cite{Gerard} and we use the formula in \cite{Ball}
to get the transition matrix element $M_{12}$. As pointed out in
\cite{Ball}, $B_s-\bar{B}_s$ mixing is very sensitive to the
combination $\delta_{LL}^d \delta_{RR}^d $ and $\delta_{LR}^d
\delta_{RL}^d$, and with the current measurement of
$B_s-\bar{B}_s$ mixing\cite{D0}, it can put rather severe
constraints on $\delta^d$s.

\em{4. $\delta M_W$ and $\delta \sin^2 \theta_{eff}:$~~} \rm In
the MSSM the corrections to $M_W$ and $\sin^2 \theta_{eff} $ are
dominated by squark loops \footnote{Slepton contribution to
$\delta M_W$ and $\delta \sin^2 \theta_{eff}$ is not important
partially because slepton is $SU(3)$ color singlet, and partially
because, due to absence of large $\tilde{l}_L-\tilde{l}_R$ mixing,
the slepton $SU(2)$ doublet $(\tilde{\nu}, \tilde{l})$ are nearly
degenerate, hence their contribution to $\delta \rho$ tends to
vanish\cite{rho}.}.
In our calculations we used the complete formula in \cite{rho} for the
corrections. We checked that, to a good approximation (at the
level of a few percent), they can be determined from the $\delta
\rho$ parameter \cite{hollik}
\begin{eqnarray}
\delta M_W \simeq \frac{M_W}{2} \frac{c_W^2}{c_W^2 -s_W^2} \delta \rho, \\
\delta \sin^2 \theta_{eff} \simeq -\frac{c_W^2 s_W^2}{c_W^2
-s_W^2} \delta \rho ,
\end{eqnarray}
where
\begin{eqnarray}
 \delta \rho=\frac{\Sigma_Z(0)}{M_Z^2} - \frac{\Sigma_W(0)}{M_W^2}.
\end{eqnarray}
Although our results are not sensitive to the soft mass parameters
of the first-generation squarks, we stress the necessity to
consider all three generations of squarks in the calculations. The
reason is the calculation of the squark loops in $W$-boson
self-energy involves the couplings of $W \tilde u_i \tilde d_j$,
which are given by $U_{\tilde u}^\dag V_{CKM} U_{\tilde d}$ with
$U_{\tilde u}$ and $U_{\tilde d}$ defined in eq.(\ref{rotation}). 
Only by considering all three
generations of squarks can one get exact UV-convergent results and
implement the SU(2) relation in eq.(\ref{SU2}) at the same time.
In \cite{hollik} the authors only considered two generations of
squarks (ignored the first generation) and by introducing an
unphysical $2 \times 2$ unitary matrix they kept their results
free of UV-divergence. We checked numerically that such an
approximation is not so good. For example, with the same
parameters for the lowest curve of Fig.8 in \cite{hollik}, our
result is $\delta M_W =11$ MeV for $\delta_{LL} = 0.6 $, smaller
than $40$ MeV obtained in \cite{hollik}. The main reason for such
sizable difference is due to the difference of  $W \tilde u_i
\tilde d_j$ couplings. The matrices $U_{\tilde u}$ and $U_{\tilde
d}$ of [6] are different from ours since they did not consider the
SU(2) relation between up- and down-squark mass matrices, and the
matrix $V_{CKM}$ of [6] is also different from ours since we used
the exact $3 \times 3$ CKM matrix while they used an approximated
$2 \times 2$ unitary matrix. Due to the strong cancellation
between different Feynman diagrams in the calculation of $\delta \rho$,
the small difference of $W \tilde u_i \tilde d_j$ couplings may lead to
sizable difference in the final result.
\vspace*{0.2cm}

 \em{5. Higgs boson mass:~~} \rm In the MSSM the
loop-corrected lightest Higgs boson mass $m_h$ is defined as the
pole of the corrected propagator matrix, which can be obtained by
solving the equation \cite{Dabelstein}
\begin{eqnarray}
 & & \left[p^2 - m_{h, tree}^2 + \hat{\Sigma}_{hh}(p^2) \right]\nonumber \\
&& \times \left[p^2 - m_{H, tree}^2 + \hat{\Sigma}_{HH}(p^2) \right]
-\left[\hat{\Sigma}_{hH}(p^2)\right]^2 = 0 ,
\label{mass-equation}
\end{eqnarray}
where $m_{h, tree}$ and $m_{H, tree}$ are the tree-level masses of
$h$ and $H$. To obtain the renormalized self-energies
$\hat\Sigma_i (p^2)$, one needs to calculate Higgs boson
self-energy and tadpole diagrams and then organize the results by
eq.(3.1) in \cite{Dabelstein}. Due to the large top-quark Yukawa
couplings, the contribution from top and stop loops is far
dominant among all SUSY contributions.  In the presence of flavor
mixings in the up-suqark mass matrix, stops will mix with other
squarks, and in this case, the dominant contribution comes from
up-squark sector. In our calculation of $m_h$, we take into
account this part of contribution as in \cite{hollik}. For
$\delta_{LR, RL}=0$, we can reproduce the results in
\cite{hollik}, but our results include $\delta_{LR, RL} \neq 0$
case.

With the parametrization of the squark mass matrices in
eqs.(\ref{up squark}$\sim$\ref{SU2}) and for small flavor mixing
parameters of down-squarks, the
quantities $\delta M_W$, $\delta \sin^2 \theta_{eff}$ and $m_h$
are sensitive to the soft masses $M_{Q_{2,3}}$ and $M_{U_{2,3}}$,
the trilinear coupling $A_t$ as well as $\delta_{LR,RL}$.
And in some cases, they are also sensitive to $\mu$, the CP-odd Higgs boson
mass $m_A$ and $\delta_{LL,RR}$.
Unlike $B_s-\bar B_s$ mixing and $b \to s \gamma$, these quantities are not
sensitive to gluino mass.
\vspace*{0.4cm}

\bf{Numerical results~~} \rm To numerically illustrate the
constraints on scharm-stop mixing parameters $\delta_{LL}$ and
$\delta_{LR}$, we need to fix other involved parameters. Here, as
an illustration, we consider the so-called $m_h^{max}$ scenario,
in which $m_h$ can be maximized \cite{benchmark}. In this
scenario, all the soft mass parameters are assumed to be
degenerate
\begin{eqnarray}
M_{SUSY} = M_{Q_i} = M_{U_i}=M_{D_i} ,
\end{eqnarray}
and the trilinear couplings are also assumed to be degenerate
$A_{u_i} = A_{d_i}$ with $(A_{u_i} - \mu \cot \beta)/M_{SUSY} =2$.
Other SUSY parameters are fixed as $\tan\beta=10$, $\mu = M_2=m_A
= m_{\tilde{g}}= 300 {\rm ~GeV}$ and $\delta_{RL}=\delta_{RR}\delta_{LR}^d=\delta_{RL}^d = \delta_{RR}^d = 0$. All the SM
parameters involved in our calculations like $M_Z$ are taken from
the Particle Data Book \cite{pdg}.

\begin{center}
\begin{figure}[tb]
%\hspace*{-1.2cm}
\epsfig{file=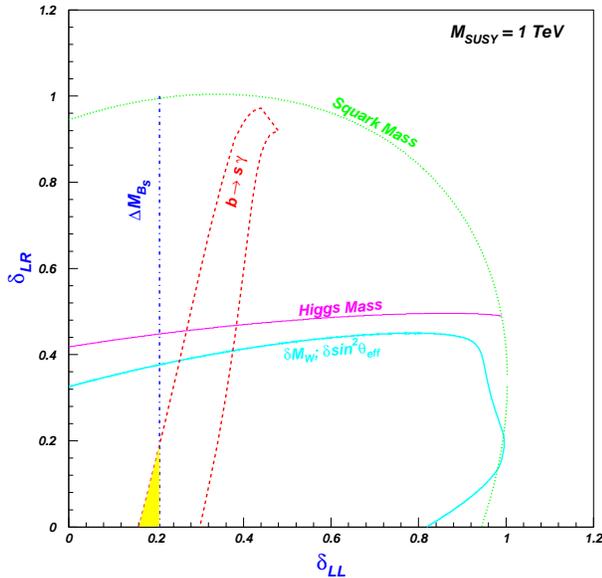,width=8cm, height=7.7cm}
\caption{The shaded area is allowed by all constraints.
The dashed-line enclosed area is allowed by $b \to s \gamma$.
For other individual constraints, the region under
or left to each curve is the corresponding allowed region.
\label{allowed}}
\end{figure}
\end{center}

We show in Fig.\ref{allowed} the constraints on the mixing
parameters $\delta_{LL}$ and $\delta_{LR}$. Here we used the LEP
experimental bounds on squark mass  \cite{pdg} and Higgs boson
mass \cite{lep-higgs}
\begin{eqnarray}
m_{\tilde{u}} > 95.7 {\rm ~GeV},~m_{\tilde{d}} > 89  {\rm~GeV},
~m_h > 92.8  {\rm ~GeV}, \label{sq bounds}
\end{eqnarray}
and required the MSSM loop contributions to $M_W$ and
$\sin^2\theta_{eff}$ not exceeding the present experimental
uncertainties \cite{lep}
\begin{eqnarray}
\Delta M_W < 34  {\rm ~MeV}, \quad \Delta \sin^2\theta_{eff} < 15 \times 10^{-5} .
\end{eqnarray}
We also used the current measurement of $ b \to s \gamma$ at $3
\sigma$ level \cite{heavy flavor}
\begin{eqnarray}
2.53 \times 10^{-4} < Br( b \to s \gamma ) < 4.34 \times 10^{-4},
\end{eqnarray}
and current favored region for $B_s-\bar{B}_s$ mixing\cite{Bs
bound,D0}
\begin{eqnarray}
0.55 < | 1 +  R | < 1.37 \label{Bs bounds}
\end{eqnarray}
where $R = M_{12}^{SUSY}/M_{12}^{SM}$.
 %%%%%%%%%%%%%%%%%%%%%%%%%%%%%%%%%%%%%%%%%%%%%%%%%%%%%%%%%%%%%%%%%%%%%%%%%%%%%%%%%%%
%%%%%%%%%%%%%%%%%%%%%%%%%%%%%%%%%%%%%%%%%%%%%%%%%%%%%%%%%%%%%%%%%%%%%%%%%%%%%%%%%%%
From Fig.\ref{allowed} we see that although the constraint
from squark mass is rather weak, the
combined constraints from $b \to s \gamma$, $B_s-\bar{B}_s$
mixing, $\delta M_W$, $\delta \sin^2 \theta_{eff}$ and $m_h$ are
quite strong and only a small area in $\delta_{LL}-\delta_{LR}$
plane survives. We make following explanations about our results:
\begin{itemize}

\item[(1)] With our fixed parameters, the charged Higgs and chargino
contributions enhance the SM prediction of $b \to s \gamma$ and thus
a none-zero gluino contribution is needed to cancel the effects.
Therefore, the $b \to s \gamma$ allowed region is an enclosed one
in Fig.\ref{allowed}

\item[(2)] In the case we considered, $\delta M_W$ and $\delta \sin^2
\theta_{eff}$ require a small $\delta_{LR}$. The reason is that
the large splitting between $\delta_{LR}$ and $\delta_{LR}^d$
can spoil the custodial symmetry between squark
doublet and hence enhance the value of $\delta \rho$. In our
calculations we fixed $\delta_{LR}^d=0$, but we checked that the
maximal allowed value of $\delta_{LR}$ is not sensitive to
$\delta_{LR}^d$ if $\delta_{LR}^d < 0.4$. For example, when
$\delta_{LR}^d$ varies from 0 to 0.4, the maximal value of
$\delta_{LR}$ only increases by 0.05.
Note that compared with $\delta_{LR}$, $\delta M_W$ and $\delta \sin^2
\theta_{eff}$ are less sensitive to $\delta_{LL}$.

\item[(3)] The Higgs mass $m_h$ requires a small $\delta_{LR}$
because $\delta_{LR}$ comes from $H_2 \tilde{c}^\ast_L
\tilde{t}_R$ interaction which can lower the value of $m_h$ via
squark loops.

\item[(4)] With our fixed parameters, $B_s-\bar{B}_s$ mixing and $
b \to s \gamma $ give the stronger constraint on $\delta_{LR}$
than $\delta \rho$ and $m_h$, but in other cases, $\delta \rho$ or
$m_h$ may give the strongest constraints. For example, in
$m_h^{max}$ scenario with $M_{SUSY} =2$ TeV, the Higgs mass
constraint becomes the strongest.
\end{itemize}

Note that in the above we only show the constraints on
$\delta_{LL}$ and $\delta_{LR}$.
Now we take a look at $\delta_{RL}$ and $\delta_{RR}$.
Since $B_s-\bar B_s$ mixing and
$b \to s \gamma$ are  not sensitive to $\delta_{RL}$ and $\delta_{RR}$,
the constraints are then only from $\delta\rho$ and $m_h$.
We found that $\delta \rho$ requires a small $\delta_{RL}$
(the constraint is similar to $\delta_{LR}$), but is insensitive
to $\delta_{RR}$.
Just like  $\delta_{LR}$, a small $\delta_{RL}$ is also
required by the Higgs mass $m_h$.
Therefore,  $\delta_{RL}$ is constrained by $\delta\rho$ and $m_h$
just like  $\delta_{LR}$; while $\delta_{RR}$ is very weakly constrained.
\vspace*{0.2cm}

\bf{Implication in top-quark FCNC process:~~} \rm With the
constraints discussed above, we examined various top-quark FCNC
decay and production processes at the LHC, some of which were
intensively studied in the literature
\cite{tcv-mssm,tch-mssm,tc-production-mssm}. We just considered
the dominant SUSY-QCD contributions to these processes. The
detailed calculations, the results and the discussions of
observability are quite lengthy and will be presented in another
paper. Here, as an illustration of the effects of the constraints
discussed in this paper, we only show in Table 1 the maximal
results predicted by the MSSM. These maximal results are obtained
by scanning the relevant SUSY parameters in the ranges
\begin{eqnarray}
&& 2 < \tan\beta < 60, \quad  0< M_{Q_i,U_i, D_i}<1 {\rm ~TeV}, \nonumber \\
&& 94 {\rm ~GeV} < m_A <1 {\rm ~TeV}, \quad 195 {\rm ~GeV} <m_{\tilde g} < 1 {\rm ~TeV},
                                                                \nonumber \\
&& 0< \delta_{LL,LR} < 1, \quad -1 {\rm ~TeV}  < \mu, M_2 < 1 {\rm ~TeV}, \nonumber \\
&& 0< \delta_{LR}^b < 0.1, \quad -2 {\rm ~TeV} < A_{t,b} < 2 {\rm
~TeV} .
\end{eqnarray}
We show in Table I two kinds of predictions: one is by requiring
the squark, chargino and neutralino masses satisfy their current
lower bounds; the other is by imposing all constraints considered
in this paper. We consider two cases: (I) only $\delta_{LL}\neq 0$
and (II) only $\delta_{LR} \neq 0$. From Table I we see that the
combined constraints can significantly decrease the MSSM
predictions of top-quark FCNC processes at the LHC.

Note that in Table I we only illustrate the cases of
$\delta_{LL}\neq 0$ and $\delta_{LR} \neq 0$.
For $\delta_{RL}\neq 0$, we found that the maximum rate of
$t \to cg$ is $1.3 \times 10^{-4}$ with only the squark mass constraints
and changed to $6 \times 10^{-5}$ with all constraints.
For $\delta_{RR} \neq 0$,  the maximum rate of $t \to cg$ is
$5.0 \times 10^{-5}$ with only the squark mass constraints
and changed to $4.85 \times 10^{-5}$ with all constraints.
This can be easily understood since, as discussed ealier, $\delta_{RR}$
is very weakly constrained and $\delta_{RL}$ is constrained by $\delta\rho$
and $m_h$.

\vspace*{0.2cm}
%%%%%%%%%%%%%%%%%%%%%%%%%%%%%%%%%%%%%%%%%%%%%%%%%%%%%%%%%%%%%%%%%%%%%%%%%%%%%%%%%%
\noindent {\small Table 1: Maximal predictions for top-quark FCNC
processes
     induced by stop-scharm mixings via gluino-squark loops in the MSSM.
     For the productions we show the hadronic cross sections at the LHC
     and include the corresponding charge-conjugate channels.
     For the decays we show the branching ratios. }
\vspace*{-0.4cm}
\begin{center}
\begin{tabular}{|c|c|c|c|c|} \hline
& \multicolumn{2}{c|}{$\delta_{LL}\neq
0$}&\multicolumn{2}{c|}{$\delta_{LR}\neq 0$} \\ \cline{2-5}
        & constraints& constraints& constraints & constraints \\
        &  masses&  all    & masses &   all      \\ \hline
$cg \to t$ & 1450 fb & 225 fb& 3850 fb & 950 fb \\ \hline $gg \to
t\bar{c}$ & 1400 fb & 240 fb & 2650 fb & 700 fb \\ \hline $ c g
\to tg$ & 800 fb & 85 fb & 1750 fb & 520 fb      \\ \hline $cg \to
t \gamma$ & 4 fb & 0.4 fb& 8 fb & 1.8 fb    \\ \hline $cg \to tZ$
& 11 fb & 1.5 fb & 17 fb & 5.7 fb           \\ \hline $c g \to t
h$ & 550 fb & 18 fb & 12000 fb & 24 fb       \\ \hline \hline $t
\to ch$ & $1.2 \times 10^{-3} $ & $ 2.0 \times 10^{-5}$ & $2.5
\times 10^{-2}$ & $ 6.0 \times 10^{-5}$ \\ \hline $t \to c g $ &
$5.0 \times 10^{-5}$ & $5.0 \times 10^{-6} $ & $1.3 \times
10^{-4}$ & $3.2 \times 10^{-5}$ \\ \hline $t \to c Z$ & $5.0
\times 10^{-6}$ & $5.7 \times 10^{-7}$ &$1.2 \times 10^{-5}$ &
$1.8 \times 10^{-6}$ \\ \hline $t \to c \gamma $ & $9.0 \times
10^{-7}$ & $1.5 \times 10^{-7}$ & $1.3 \times 10^{-6}$ & $5.2
\times 10^{-7}$ \\ \hline
\end{tabular}
\end{center}
%%%%%%%%%%%%%%%%%%%%%%%%%%%%%%%%%

\vspace*{0.2cm} \bf{Conclusion:~~} \rm We examined current
experimental constraints on scharm-stop flavor mixing in the MSSM,
which arise from the experimental bounds on squark and Higgs boson
masses, the precision measurements of the $\delta\rho$ parameter,
as well as the experimental data on $B_s-\bar{B}_s$ mixing and $b
\to s \gamma$. We found that the combined analysis of these
constraints can severely constrain the $\tilde{c}_L-\tilde{t}_L$
and $\tilde{c}_L-\tilde{t}_R$ mixings; while the
$\tilde{c}_R-\tilde{t}_L$ mixing is constrained only by
$\delta\rho$ and Higgs mass, and the $\tilde{c}_R-\tilde{t}_R$
mixing can almost elude any constraints. Such constraints can
significantly alter the predictions of the top-quark FCNC
processes induced by scharm-stop mixings at the LHC and thus
should be taken into account in the study of observability of
these top-quark FCNC processes. \vspace*{0.2cm}
\vspace*{0.2cm}

\bf{Acknowlegement:~~} \rm
This work is supported in part by the Israel Science Foundation,
the Fund for the Promotion of Research at Technion, the
Grant-in-Aid for Scientific Research (No. 14046201) from the Japan
Ministry of Education, Culture, Sports, Science and Technology and
by the National Natural Science Foundation of China under Grant
No. 10475107, 10505007 and 10375017.

\end{document}